\documentclass[sigconf]{acmart}

\usepackage{multirow}
\usepackage{subcaption}

\AtBeginDocument{%
  }

\setcopyright{acmlicensed}
\copyrightyear{2018}
\acmYear{2018}
\acmDOI{XXXXXXX.XXXXXXX}
\acmConference[Conference acronym 'XX]{Make sure to enter the correct
  conference title from your rights confirmation email}{June 03--05,
  2018}{Woodstock, NY}
\acmISBN{978-1-4503-XXXX-X/2018/06}

\usepackage{xcolor}
\usepackage[normalem]{ulem} 

\begin{document}

\title{HIT Model: A Hierarchical Interaction-Enhanced Two-Tower Model for Pre-Ranking Systems}
\author{Haoqiang Yang}
\authornote{Both authors contributed equally to this research.}
\affiliation{%
  \institution{Tencent}
  \city{Shenzhen}
  \state{Guangdong}
  \country{China}
}
\email{yanghaoqiang1998@gmail.com}

\author{Congde Yuan}
\authornotemark[1]
\affiliation{%
  \institution{Tencent}
  \city{Shenzhen}
  \state{Guangdong}
  \country{China}
}
\email{congdeyuan@gmail.com}

\author{Kun Bai}
\affiliation{%
  \institution{Tencent}
  \city{Shenzhen}
  \state{Guangdong}
  \country{China}}
\email{bookerbai@tencent.com}

\author{Mengzhuo Guo}
\authornote{Corresponding author.}
\affiliation{%
  \institution{Sichuan University}
  \city{Chengdu}
  \state{Sichuan}
  \country{China}
}
\email{mengzhguo@scu.edu.cn}

\author{Wei Yang}
\affiliation{%
 \institution{Tencent}
  \city{Shenzhen}
  \state{Guangdong}
  \country{China}}
 \email{viviwyang@tencent.com}

\author{Chao Zhou}
\affiliation{%
  \institution{Tencent}
  \city{Shenzhen}
  \state{Guangdong}
  \country{China}}
  \email{derekczhou@tencent.com}

\renewcommand{\shortauthors}{Trovato et al.}

\begin{abstract}
Online display advertising platforms rely on pre-ranking systems to efficiently filter and prioritize candidate ads from large corpora, balancing relevance to users with strict computational constraints. The prevailing two-tower architecture, though highly efficient due to its decoupled design and pre-caching, suffers from cross-domain interaction and coarse similarity metrics, undermining its capacity to model complex user-ad relationships. 
In this study, we propose the Hierarchical Interaction-Enhanced Two-Tower (HIT) model, a new architecture that augments the two-tower paradigm with two key components: \textit{generators} that pre-generate holistic vectors incorporating coarse-grained user-ad interactions through a dual-generator framework with a cosine-similarity-based generation loss as the training objective,  
and \textit{multi-head representers} that project embeddings into multiple latent subspaces to capture fine-grained, multi-faceted user interests and multi-dimensional ad attributes. 
This design enhances modeling effectiveness without compromising inference efficiency. Extensive experiments on public datasets and large-scale online A/B testing on Tencent's advertising platform demonstrate that HIT significantly outperforms several baselines in relevance metrics, yielding a 1.66\% increase in Gross Merchandise Volume and a 1.55\% improvement in Return on Investment, alongside similar serving latency to the vanilla two-tower models. The HIT model has been successfully deployed in Tencent's online display advertising system, serving billions of impressions daily. The code is available at \url{https://github.com/HarveyYang123/HIT_model}.

\end{abstract}





\maketitle

\section{Introduction}

\begin{figure}[!htbp]
    \centering
    \includegraphics[trim=20 0 20 0 clip, width=0.4\textwidth]{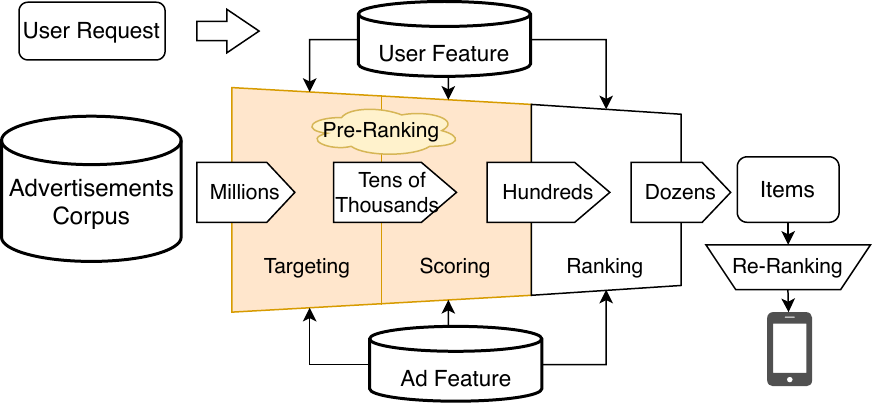}
    \caption{Online display advertising system.}
    \label{ad_flow}
\end{figure}

Online display advertising systems serve as a critical revenue stream for numerous digital platforms, including Google \citep{yi2019sampling, gui2015truth}, Meta \citep{zhang2024scaling, wiese2020framework}, Alibaba \citep{jiang2024gats, wei2022intelligent, ouyang2021learning}, and Tencent \citep{zhang2024deep, zheng2022hien, yuan2022actor, guo2024bayesian}. 
These systems recommend the most interesting advertisements to users and facilitate them to purchase the products, thereby optimizing advertiser profits \citep{zhao2021dear, guo2024bayesian}.
Typically organized in a cascaded architecture, as depicted in Figure \ref{ad_flow}, these systems commence with \textit{targeting}, which identifies a candidate set of ad impressions from a vast corpus comprising millions of ads. This candidate set is subsequently \textit{scored} to distill several hundred most relevant ads. Finally, the \textit{ranking} and \textit{re-ranking} stages refine this selection to a handful of ads, completing the pipeline.

The \textit{pre-ranking} system, encompassing the targeting and scoring tasks highlighted in the shaded area of Figure \ref{ad_flow}, plays a crucial role in the architecture of online display advertising systems. This phase is critical as it precedes the evaluation and final display of ads \citep{xu2022mixture}. Its primary objective is to ensure that the ad impressions are not only relevant to the targeted users, thereby enhancing \textit{model effectiveness}, but also that they are processed in a manner that is not computationally burdensome, ensuring \textit{inference efficiency}. Consequently, the principal challenge lies in achieving an optimal balance between these two tasks \citep{li2022inttower}.

Due to the stringent requirement that pre-ranking systems in large-scale advertising platforms must filter millions of ad candidates per user within milliseconds, the standard \textit{two-tower model} has become the de facto in industry \citep{huang2013learning, li2022virt, wang2021distilled}. While this architecture excels in computational efficiency by pre-caching techniques, its effectiveness is constrained by (1) \textbf{a lack of coarse-grained information exchange with the opposite tower, preventing cross-domain dependency modeling,} and (2) \textbf{a simple dot-product-based similarity computation, failing to capture fine-grained semantic relationships between user and ad embeddings.} Therefore, traditional two-tower-based models may miss the critical multi-faceted user interests and multi-dimensional ad attributes essential for precise matching.

To address these two limitations, we propose a \textbf{H}ierarchical \textbf{I}nteraction-Enhanced \textbf{T}wo-Tower (\textbf{HIT}) model, as illustrated in Figure \ref{hit_model}. The proposed HIT model is comprised of two new components: the \textbf{\textit{generators}} and \textbf{\textit{multi-head representers}}. The generators establish a holistic representation of user/ad static features, enabling the early generation of embeddings to mimic the coarse-grained information from the counterpart tower before the multi-head representers. To this end, we adopt a cosine-similarity-based generation loss and separately learn positive and negative samples by two generators. Such a dual-generator approach treats positive and negative samples differently since they contain rich information about the targeted and non-targeted user/ad.

The multi-head representers employ linear projections to map the generated user/ad embeddings into different latent sub-spaces. To effectively extract user/ad representations, we first capture the strongest signal from each vector of the user's multi-faceted interest (or the ad's multi-dimensional attributes) to determine the most relevant user/ad interactions, and then aggregate these interactions to obtain a final score. 
It accounts for the fine-grained multi-faceted user interests and multi-dimensional ad attributes, yielding more precise matching scores. Moreover, the multi-head representers retain the advantage of pre-cached ad embeddings, thereby maintaining high serving efficiency.

Our contributions can be summarized as follows. First, we propose the HIT model for pre-ranking systems, which addresses the issue of insufficient information interaction while preserving the serving efficiency of the two-tower paradigm. The HIT model contains two new components: the generators for pre-generating coarse-grained vectors of user/ad cross-domain feature fusions and multi-head representers for projecting embeddings into multi-faceted user interests (or multi-dimensional ad attributes). Second, we find that using a single vector to embed user/ad characteristics, as stated in traditional two-tower models, is insufficient to capture user/ad granular characteristics. The proposed generators employ a cosine-similarity-based loss to eliminate vector magnitude discrepancies for more accurate coarse-grained generation vectors, while the multi-head representers adopt a max-then-sum operation to accomplish fine-grained matching between multi-faceted user interests and multi-dimensional ad attributes. Finally, comprehensive evaluations of public datasets demonstrate that the HIT model consistently outperforms baseline models. Rigorous online A/B tests on Tencent's online display advertising platform further confirm its effectiveness, with significant gains in GMV, ROI, and reduced response times. The HIT model has been successfully deployed on a large-scale advertising platform, impacting billions of daily impressions.

\section{Related Work}
\label{sec-related}
\begin{figure*}[!htbp]
    
\centering
    \begin{subfigure}{.22\textwidth}
    \centering
\includegraphics[trim=300 10 300 10, scale=0.5]{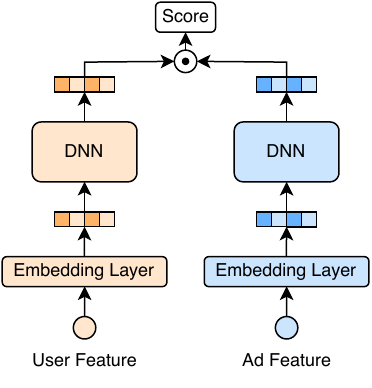}
    \subcaption{Vanilla two-tower.\label{fig-vanillla}}
    \end{subfigure}
    \hfill
    \begin{subfigure}{.22\textwidth}
    \centering
\includegraphics[trim=300 10 300 10, scale=0.5]{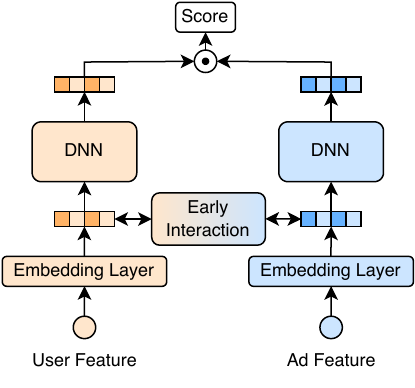}
    \subcaption{Early interaction.\label{fig-early}}
    \end{subfigure}
    \hfill
    \begin{subfigure}{.22\textwidth}
    \centering
\includegraphics[trim=300 10 300 10, scale=0.5]{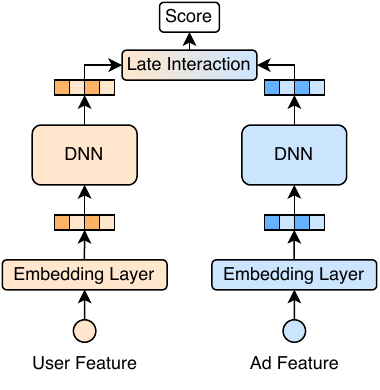}
    \subcaption{Late interaction.\label{fig-late}}
    \end{subfigure}
    \hfill
    \begin{subfigure}{.22\textwidth}
    \centering
\includegraphics[trim=300 10 300 10, scale=0.5]{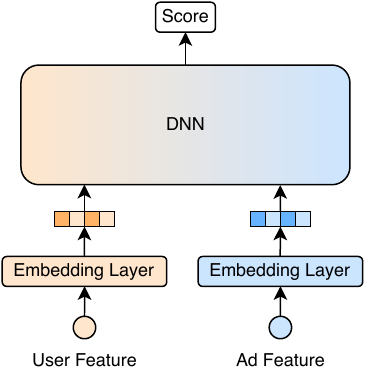}
    \subcaption{All-to-all interaction.\label{fig-alltoall}}
    \end{subfigure}
    
    \caption{Illustrations of four types of two-tower model structures.}
    \label{all_interaction_model}
\end{figure*}

The two-tower architecture has emerged as a dominant paradigm in pre-ranking systems, widely employed in online display advertising and recommender systems. As pre-ranking systems operate at the forefront of the decision pipeline, their efficiency and effectiveness significantly impact subsequent stages. The two-tower model leverages two independent neural networks to encode users and ads, thereby enabling joint training while also allowing each tower to be pre-trained offline to minimize computational overhead during online inference. This design not only ensures scalability but also achieves lower latency in prediction compared to traditional shallow learning methods, such as logistic regression \citep{mcmahan2013ad} and gradient boosting decision trees \citep{wang2016mobile}. Consequently, the two-tower model has become a cornerstone in modern recommender and advertising systems \citep{huang2013learning, li2022inttower}. We categorize it into four mainstream approaches based on their interaction mechanisms: \textbf{vanilla}, \textbf{early interaction}, \textbf{late interaction}, and \textbf{all-to-all interaction} models, as illustrated in Figure \ref{all_interaction_model}.


\textbf{Vanilla Two-tower Model.} The vanilla two-tower model, exemplified by DSSM \citep{huang2013learning}, encodes user and ad features independently without explicit interaction between the two sides. This simplicity enables efficient offline graph construction and nearest-neighbor retrieval, ensuring low-latency online inference. However, as the complexity of user and ad features increases, the absence of explicit feature interactions limits the model's ability to fully capture underlying relationships, resulting in suboptimal performance compared to more advanced methods.


\textbf{Early Interaction Model.} This strategy involves feature interaction after the embedding layers but before the encoding modules (Figure \ref{fig-early}). For instance, \citet{yu2021dual} proposes a Dual Augmented Two-tower model (DAT) that learns auxiliary vectors for users and ads based on ads and user embedding vectors, respectively. Such a process is similar to ``distillation'' that keeps the information most appropriate for describing the user-ad interactions \citep{li2022virt, wang2021distilled, lu2022ernie}. Another recent research, the mixture of virtual-kernel experts (MVKE), has modeled user-ad interactions through a multi-task learning scheme to learn user interests in different ads \citep{xu2022mixture}.


\textbf{Late Interaction Model.} Instead of modeling feature interactions before the encoding module, the late interaction directly impacts the encoded users and ads vectors, as shown in Figure \ref{fig-late}. Compared to early interaction models, this type incorporates deeper interactions between users and ads since it works after transforming the embedded vectors through shallow Deep Neural Networks (DNNs) \citep{qu2016product}. For example, to accelerate the computational efficiency, \citet{humeau2019poly} developed the poly-encoder to learn cross-features through a final attention mechanism structure. \citet{li2022inttower} propose a lightweight Multi-Layer Perceptron (MLP), termed the IntTower, which is introduced to model complex cross features between users and ads.


\textbf{All-to-all Interaction Model.} The all-to-all interaction utilizes an MLP to directly model the user-ad feature interactions right after the embedding layers (Figure \ref{fig-alltoall}). This approach departs from the paradigm that separately encodes the embedded user and ad vectors, thus overcoming the limitations of inner product representations in two-tower models. In this way, the advanced model structures, such as Wide \& Deep \citep{cheng2016wide}, Deep \& Cross Networks (DCN) \citep{wang2017deep}, AutoInt \citep{song2019autoint}, and COLD \citep{wang2020cold}, can be used to account for high-order and extremely complex user-ad interactions, thus enhancing the model accuracy.  
However, ad vectors cannot be pre-computed and stored with all-to-all interaction, significantly slowing inference speed.


In summary, vanilla two-tower models are pioneers in balancing inference efficiency and computational effectiveness. Early and late interaction models extend the two-tower framework to enhance feature interactions, improving model effectiveness. In contrast, all-to-all interaction models completely abandon the two-tower structure for more thorough feature interactions, sacrificing inference efficiency for greater model effectiveness. Our proposed HIT model retains the two-tower structure, establishing latent correlations between users and ads, and fully leveraging the rich and diverse characteristics of users and ads. It better aligns with real-world data distributions, resulting in significant performance improvements that surpass those of all-to-all interaction models without compromising inference efficiency.


\section{Preliminaries}


\subsection{Problem Definition}




In online display advertising, when a user request is received, the system combines user features $\textbf{x}_u$ (static feature $\textbf{x}_{ub}$: age, gender, occupation; dynamic feature $\textbf{x}_{ug}$: location, active period) and ad features $\textbf{x}_a$ (static feature $\textbf{x}_{ab}$: category, tag; dynamic feature $\textbf{x}_{ag}$: click count) to compute scores for all eligible ads via $s(\textbf{x}_u, \textbf{x}_a)$, where $s(\cdot)$ is a DNN trained on historical data to predict user click probabilities. A top-k truncation step then selects the most user-preferred ads based on these scores.

\subsection{Basic Structure}
The basic structure of a vanilla two-tower model, as shown in the "two-tower backbone" part (in green) in Figure \ref{hit_model}, consists of two distinct components: a user tower and an ad tower, along with a scoring and online serving module.



\subsubsection{User Tower.}
The user features $\textbf{x}_{ub}$ and $\textbf{x}_{ug}$ are first passed through an embedding layer, converting the sparse features into dense vectors $\mathbf{e}_{ub} \in \mathbb{R}^{d \times n_{ub}}$ and $\mathbf{e}_{ug} \in \mathbb{R}^{d \times n_{ug}}$, where $d$ denotes the embedding dimension, $n_{ub}$ and $n_{ug}$ are the numbers of user static and dynamic features, respectively. These dense vectors are subsequently concatenated to obtain $\mathbf{e}_{u} = [\mathbf{e}_{ub}, \mathbf{e}_{ug}] \in \mathbb{R}^{d \times (n_{ub}+n_{ug})}$, which is then fed into a DNN with $L$ layers:
\begin{equation}
  \mathbf{h}^{(l)} = \text{ReLu}(\mathbf{W}^{(l)}\mathbf{h}^{(l-1)} + \mathbf{b}^{(l)}), \; \text{for } l=1,2,\ldots,L,
    \label{eq:hidden_layer}
\end{equation}
where $\mathbf{W}^{(l)} \in \mathbb{R}^{d_{l} \times d_{l-1}}$, $\mathbf{b}^{(l)} \in \mathbb{R}^{d_{l}}, \mathbf{h}^{(l)} \in \mathbb{R}^{d_{l}}$ are the weight matrix, bias vector, and output of the $l$-th layer, with $d_l$ being the layer width, and $\mathbf{h}^{(0)} = \mathbf{e}_{u}, d_0 = d \times (n_{ub}+n_{ug})$. The final user embedding is obtained through $L_2$ norm: $\mathbf{h}_u = \left \Vert \mathbf{h}^{(L)} \right \Vert_2$.

\subsubsection{Ad Tower.}

Similarly to the user tower, the ad features $\textbf{x}_{ab}$ and $\textbf{x}_{ag}$ are converted into $\mathbf{e}_{ab} \in \mathbb{R}^{d \times n_{ab}}$ and $\mathbf{e}_{ag} \in \mathbb{R}^{d \times n_{ag}}$, where $n_{ab}$ and $n_{ag}$ denote the numbers of ad static and dynamic features. These vectors are concatenated to form $\mathbf{e}_{a} = [\mathbf{e}_{ab}, \mathbf{e}_{ag}] \in \mathbb{R}^{d \times (n_{ab}+n_{ag})}$, and are fed into a DNN. The final ad embedding $\mathbf{h}_a$ is obtained by applying $L_2$ norm to the output of the last layer.

\subsubsection{Combining Embedded User and Ad Vectors.} After the user embedding and ad embedding are computed, a score $\hat{y}$ is obtained by inner product: 
\begin{equation}
\hat{y} = \mathbf{h}_u^T \mathbf{h}_a.
\end{equation}

In the pre-ranking system, each training sample is constituted in the form of $(\textbf{x}_{ub}, \textbf{x}_{ug},\textbf{ x}_{ab}, \textbf{x}_{ag}, y)$ with $y \in \{0,1\}$ serving as the label, where 1 indicates a positive signal (the user likes the ad), and 0 otherwise. The model parameters can be optimized by minimizing the discrepancy between output $\hat{y}$ and the label $y$.

In online systems, ad embeddings are pre-computed and cached for efficient retrieval \citep{huang2013learning}, therefore only the requested user embedding need to be calculated on the fly, which significantly improves computational efficiency. However, the lack of user-ad feature interaction leads to inaccurate predictions due to missing information in the construction of $\mathbf{h}_u$ and $\mathbf{h}_a$. Thus, we propose the HIT model.



\section{The Proposed HIT Model}

\begin{figure}[!htbp]
\centering
\includegraphics[trim=25 40 25 10 clip, width=1\columnwidth]{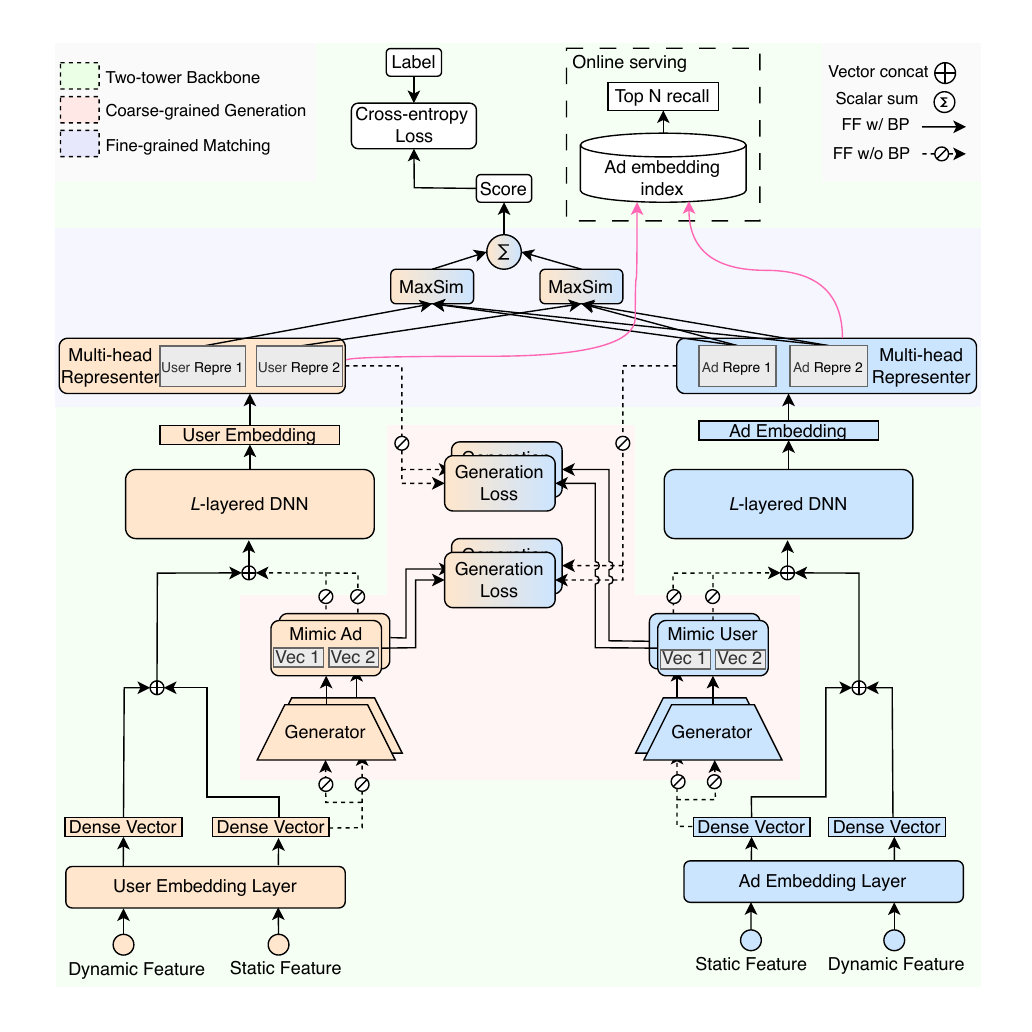}
\caption{Overview of the HIT model architecture.
}
\label{hit_model}
\end{figure}

The overall architecture of the proposed HIT model, depicted in Figure \ref{hit_model}, extends the vanilla two-tower model through the incorporation of two novel modules: \textit{coarse-grained generation} (in pink) and \textit{fine-grained matching} (in purple).


\subsection{Coarse-grained Generation}

The key component of coarse-grained generation is the \textbf{generator}, which is a simple multilayer perceptron network $g_k(\cdot)$ with two hidden layers that accepts \textit{only} static features ($\textbf{x}_{ub}$ or $\textbf{x}_{ab}$) as input and produces representations $\mathbf{r^m_\mathit{k}} \in \mathbb{R}^{p}$ manifesting information about the opposite tower. For example, the computation process of the user tower can be expressed as follows:
\begin{equation}
\mathbf{r^m_{\mathit{k}, \delta}} = g_k(\textbf{x}_{\delta b}), \quad \text{for } k=1,2,\ldots,K,
\label{eq:r_m_equation}
\end{equation}
where $\mathbf{r^m_{\mathit{k},\delta}}$ denotes the \textbf{m}imic users/ads 
(with $\delta=u/a$ representing user and ad, respectively; the subscript $\delta$ in 
$\textbf{x}_{\delta b}$ carries the same meaning)
vectors $\mathbf{r^m}$ in generator $k$, and $K$ is a hyperparameter indicating the number of generators, note that $K$ is related to the number of label types. 
For notation simplicity, we use $\mathbf{r^m_{\mathit{k}}}$ as a substitute for $\mathbf{r^m_{\mathit{k},\delta}}$, focusing on the user side. 
In this study, since the label has positive and negative types, we set $K=2$. Nevertheless, we can use smaller $K$ by feeding only positive or negative samples to the generator and greater $K$ by defining more types of labels. The training process of a generator involves comparing its output with those produced by the following multi-head representer, from which a generation loss is computed. The generation loss will be elaborated on in the following subsection.


It is noteworthy that generators simulate the high-level representations of the opposing tower, i.e., the generators in the user tower simulate ad representations, while the generators in the ad tower simulate user representations. The motivation can be intuitively explained by an example. Suppose a wealthy young woman(user) is more likely to be interested in expensive luxury brands (target ads) than cheap, low-quality goods (non-target ads). The generator in the user-tower mimics the the target ads information so that the user-tower is exposed to the ads information. To this end, the static features, describing the fundamental interests and attributes of the users and ads, are more important for a generator than the dynamic features. 
Thus, the proposed HIT model uses only static features as input for generators to fit the high-level representations of target and non-target entities, constructing correlations between static features and high-level representations through neural networks.


After obtaining the mimic representations $\mathbf{r^m_\mathit{k}}$, they need to be first normalized to unit length using L2 normalization, followed by a simple concatenation operation $f(\cdot)$ to consolidate them with the dense vector $\mathbf{e}_{u}$. 
Denote the concatenate vector as $\mathbf{e^i}$:
\begin{equation}
\mathbf{e^i} = f(\mathbf{r^m_1}, \mathbf{r^m_2}, \ldots, \mathbf{r^m_\mathit{K}}, \mathbf{e}_u).
\end{equation}
which is fed into a DNN, resulting in the user embedding $\mathbf{h^{i}_\mathit{u}} \in \mathbb{R}^{d_L}$. Similarly, for the ad tower, the ad embedding $\mathbf{h^{i}_\mathit{a}}  \in \mathbb{R}^{d_L}$ is obtained.

\subsection{Fine-grained Matching}
In this module, the user/ad embedding $\mathbf{h^{i}_\mathit{u}}$/$\mathbf{h^{i}_\mathit{a}}$ are fed into a \textbf{multi-head representer} to capture multi-dimensional user-ad interaction. In particular, the multi-head representer maps user/ad embedding to different latent sub-spaces to extract more informative representation, mirroring that a user has multi-faceted interests and an ad has multi-dimensional attributes. The projection process for the user $u$ can be described mathematically as follows:
\begin{equation}
  \mathbf{r}_{u, \mathit{(j)}} = \mathbf{W}_{u, \mathit{(j)}}\mathbf{h^{i}_\mathit{u}} + \mathbf{b}_{u, \mathit{(j)}}, \quad \text{for } j=1,2,\ldots,J,
    \label{eq:multi_head_representer_user}
\end{equation}
where $J$ is a hyperparameter indicating the number of sub-spaces mapped by the multi-head representer, $\mathbf{W}_{u,\mathit{(j)}} \in \mathbb{R}^{z \times d_{L}}$ and $\mathbf{b}_{u, \mathit{(j)}} \in \mathbb{R}^{z}$ are the transformation matrix and bias vector for the $j$-head sub-space, $\mathbf{r}_{u, \mathit{(j)}} \in \mathbb{R}^{z}$ is the high-level representation for user of $j$-head in the multi-head representer.
Similarly, the representation vector $\mathbf{r}_{a, \mathit{(j)}} \in \mathbb{R}^{z}$ for ad $a$ under $j$-head sub-space is expressed as:
\begin{equation}
  \mathbf{r}_{a, \mathit{(j)}} = \mathbf{W}_{a, \mathit{(j)}}\mathbf{h^{i}_\mathit{a}} + \mathbf{b}_{a, \mathit{(j)}}, \quad \text{for } j=1,2,\ldots,J,
    \label{eq:multi_head_representer_ad}
\end{equation}
where $\mathbf{W}_{a, \mathit{(j)}} \in \mathbb{R}^{z \times d_{L}}$ and $\mathbf{b}_{a, \mathit{(j)}} \in \mathbb{R}^{z}$  are the transform matrix and bias vector. 

We consider the matching degree between each user interest and all ad attributes in order to discover the most appropriate matching ad attribute, and then sum up all facets of interest. In this way, the final score $\hat{y}$ is obtained by:
\begin{equation}
\hat{y} = \sum_{j_u=1}^J \max_{j_a \in \{1,2,...,J\}} \left\{ (\mathbf{r}_{u, \mathit{(j_u)}})^T \mathbf{r}_{a, \mathit{(j_a)}} \right\}.
    \label{eq:score}
\end{equation}
In online serving, ad representations $\mathbf{r}_{a, \mathit{(j)}}$ are pre-computed and cached in advance, thereby retaining the computational efficiency advantage of the two-tower architecture.

\subsection{Model Optimization}
As illustrated in Figure \ref{hit_model}, the optimization of the HIT model involves two types of losses: the generation loss and the cross-entropy loss.

\subsubsection{Generation Loss.} 
The  $J$ head transformations $\mathbf{r}_{u, \mathit{(j)}}$ obtained from Eq.\eqref{eq:multi_head_representer_user} can be concatenated for parallel computing, and the result is the user representation vector $\mathbf{r}_{u} = [\mathbf{r}_{u, (1)}, \mathbf{r}_{u, (2)}, \dots, \mathbf{r}_{u, (J)}] \in \mathbb{R}^{z \times J}$ (the same process to get $\mathbf{r}_{a} \in \mathbb{R}^{z \times J}$). The generation loss $\mathcal{L}_{gu}$, which is designed to minimize the distance between $\mathbf{r}_{u}$ and $\mathbf{r^m_\mathit{k}} \in  \mathbb{R}^{p} $ derived from Eq.\eqref{eq:r_m_equation}, where $p=z \times J$, is represented as follows:
\begin{equation}
\mathcal{L}_{gu} = -\frac{1}{N}\sum_{i=1}^{N}(y_i \text{Dist}(\mathbf{r}_{u}, \mathbf{r^m_1}) + (1-y_i) \text{Dist}(\mathbf{r}_{u}, \mathbf{r^m_2})), 
    \label{eq:loss_generation}
\end{equation}
where $N$ is the total number of samples, $y_i$ is the ground truth label of $i$-th sample, and $\text{Dist}(\cdot)$ is cosine distance. Cosine distance is preferred here because it focuses on vector orientation (reflecting user/ad characteristics) rather than magnitude, aligning with the generator's objective of distinguishing distinct representations. Generator 1 ($\mathbf{r^m_1}$, target generator) and Generator 2 ($\mathbf{r^m_2}$, non-target generator) are trained on positive and negative samples, respectively. Consequently, $\mathcal{L}_{gu}$ aggregates these distances across all generators in the user tower.

The generation loss $\mathcal{L}_{ga}$ for the ad tower can be computed in an identical procedure as in Eq.\eqref{eq:loss_generation}.



\subsubsection{Cross-entropy Loss.} 
We use cross-entropy loss (CEloss) to calculate the difference between the predicted scores and the true labels, which is defined as follows:
\begin{equation}
\mathcal{L}_c  = -\frac{1}{N}\sum_{i=1}^{N}(y_i \text{log}(\sigma(\hat{y}_i)) + (1-y_i) \text{log}(1-\sigma(\hat{y}_i))),
    \label{eq:loss_cross_entropy}
\end{equation}
where $\sigma(\cdot)$ is the sigmoid function , and $\hat{y}$ is the $i$-th prediction score derived from Eq.\eqref{eq:score}.

\subsubsection{Total Loss.} 
The total loss $\mathcal{L}$ is calculated as follows:
\begin{equation}
\mathcal{L}  = \mathcal{L}_c + \alpha (\mathcal{L}_{gu}+\mathcal{L}_{ga}),
    \label{eq:loss_total}
\end{equation}
where $\alpha$ is a hyperparameter used to balance the weight of the cross-entropy loss and generation loss, as shown in Figure \ref{hit_model}, the HIT model is trained using the backpropagation algorithm in an end-to-end manner. However, to enable generators to focus on fitting high-level representations without interfering with the backbone network in the model, both the input and output of generators undergo ``stop gradient'' operations. Additionally, to ensure that the generation loss does not affect the training itself, the representation outputs of the multi-head representer are subjected to ``stop gradient'' operations within the computation of the generation loss.




\section{Experiments}

In this section, we answer the following questions: 
\\(\textit{\textbf{Q1}}) \textit{What are the HIT model's performances compared to baselines?} \\(\textit{\textbf{Q2}}) \textit{What is the impact of each component on the performance?} \\(\textit{\textbf{Q3}}) \textit{How to explain the interaction mechanism in the HIT model?}  \\(\textit{\textbf{Q4}}) \textit{What are the online performances of the HIT model?}

\subsection{Experimental Setup}
In this section, three public datasets, MovieLens, Amazon (Electro), and Alibaba, are used as offline evaluation datasets. Following \citet{huang2019fibinet}, the samples are randomly divided into two parts: 80\% for training and the remaining 20\% for testing. This ensures that our model is trained on a sufficiently large dataset while still having enough samples to evaluate its performance on unknown data. The detailed statistics of datasets are provided in Table \ref{table:statistics of datasets}.

\begin{table}[!htbp]
\centering
{\footnotesize
\begin{tabular}{ccccccc}
\toprule
Dataset & Users & Items & Samples \\
\midrule
MovieLens-1M & 6,040 & 3,952 & 1,000,000 \\
Amazon(Electro) & 192,403 & 630,001 & 1,689,188 \\
Alibaba & 1,061,768 & 785,597 & 26,557,961 \\
\bottomrule
\end{tabular}
\caption{Basic statistics of each dataset.}
\label{table:statistics of datasets}}
\end{table}


We empirically set the feature embedding dimension to 32 and the training batch size to 256. Each DNN consists of a three-layer MLP with hidden dimensions $[300, 300, 32]$. The parameters of the HIT model are temporarily fixed as: $ \alpha = 10^{-3}$ in Eq.\eqref{eq:loss_total}; $J=2$ in Eq.\eqref{eq:multi_head_representer_user} and Eq.\eqref{eq:multi_head_representer_ad}; output dimension $z=16$ for the multi-head representer. The generator is a two-layered MLP with $[64, 32]$. Other detailed implementations can be found in the open-source code. We conduct experiments with 2 Tesla T4 GPUs.


\subsection{Experimental Results Compared to Baselines}

\begin{table*}[!htbp]
\centering 
    \resizebox{\textwidth}{!}{
\begin{tabular}{lccccccccccc}
\bottomrule
 \multirow {2}{6em}{Type} & \multirow {2}{6em}{\centering{Model}} & \multirow {2}{6em}{\centering{\# params.}}& \multicolumn{3}{c}{Alibaba} & \multicolumn{3}{c}{MovieLens} & \multicolumn{3}{c}{Amazon} \\
 \cline{4-12}
 & & & AUC & CEloss & RelaImpr & AUC & CEloss & RelaImpr & AUC & CEloss & RelaImpr \\
 \cline{1-12}
\multirow {1}{6em}{Vanilla} 
& DSSM \citep{huang2013learning} & 0.74M & 0.6579 & 0.2292 & 0\% & 0.8697 & 0.4559 & 0\% & 0.8469 & 0.4313 & 0\% \\
\hline
\multirow {2}{6em}{Early} 
 & DAT \citep{yu2021dual} & 0.76M & 0.6598 & 0.2279 & 1.20\% & 0.8712 & 0.4556 & 0.40\% & 0.8480 & 0.4278 & 0.31\% \\
 & MVKE \citep{xu2022mixture} & 0.84M & 0.6625 & 0.2276 & 2.91\% & 0.8720 & 0.4511 & 0.62\% & 0.8503	& 0.4324 & 0.98\% \\
\hline
\multirow {2}{6em}{Late} 
 & Poly-Encoder \citep{humeau2019poly} & 0.87M & 0.6657 & 0.2273 & 4.94\% & 0.8734 & 0.3971 & 1.00\% & 0.8595 & 0.3855 & 3.63\% \\
 & IntTower \citep{li2022inttower} & 1.18M & 0.6827 & 0.2245 & 15.71\% & 0.8974 & 0.3128 & 7.49\% & 0.8696 & 0.3309 & 7.91\% \\
\hline
 \multirow {5}{6em}{All-to-all}
 & Wide\&Deep \citep{cheng2016wide} & 0.68M & 0.6814 & 0.2250 & 14.88\% & 0.8820 & 0.3344 & 3.32\% & 0.8615 & 0.3409 & 4.20\% \\
 & DeepFM \citep{guo2017deepfm} & 0.68M & 0.6820 & 0.2247 & 15.26\% & 0.8920 & 0.3211 & 6.03\% & 0.8643 & 0.3405 & 5.05\% \\
 & DCN \citep{wang2017deep} & 0.68M & 0.6831 & 0.2244 & 15.96\% & 0.8964 & 0.3151 & 7.22\% & 0.8665 & 0.3366 & 5.65\% \\
 & AutoInt \citep{song2019autoint} & 0.70M & 0.6867 & 0.2238 & 18.24\% & 0.8948 & 0.3192 & 6.79\% & 0.8686 & 0.3351 & 6.25\% \\
 & COLD \citep{wang2020cold} & 0.68M & 0.6816 & 0.2248 & 15.01\% & 0.8836 & 0.3297 & 3.75\% & 0.8633 & 0.3402 & 4.72\% \\
 \hline
 \multirow {1}{6em}{Hierarchical} 
 & HIT(ours) & 0.79M & \textbf{0.7226} & \textbf{0.2104} & \textbf{40.98\%} & \textbf{0.9048} & \textbf{0.3026} & \textbf{9.49\%} & \textbf{0.8784} & \textbf{0.3225} & \textbf{9.08\%}\\
\bottomrule
\end{tabular}}
\caption{Average results compared to baselines over 10 repetitions. The comparison baseline for the RelaImpr metric is the vanilla two-tower model. ($\text{RelaImpr} = \left(\frac{\text{AUC(model)} - 0.5}{\text{AUC(base)} - 0.5} - 1\right) \times 100\%$.) The improvement is significant at $\alpha=0.01$.}
\label{table:Comparative Experiment} 
\end{table*}

Table \ref{table:Comparative Experiment} answers \textit{\textbf{Q1}} and reports the average results when comparing the proposed HIT model to the following baselines: \textbf{DSSM} \citep{huang2013learning}, \textbf{DAT} \citep{yu2021dual}, \textbf{MVKE} \citep{xu2022mixture}, \textbf{Poly-Encoder} \citep{humeau2019poly}, \textbf{IntTower} \citep{li2022inttower}, \textbf{Wide\&Deep} \citep{cheng2016wide}, \textbf{DeepFM} \citep{guo2017deepfm}, \textbf{DCN} \citep{wang2017deep}, \textbf{AutoInt} \citep{song2019autoint}, and \textbf{COLD} \citep{wang2020cold}. These models are introduced in Section \ref{sec-related} and can be classified into a specific two-tower structure. We consider three metrics: \textbf{CEloss}, as shown in Eq.\eqref{eq:loss_cross_entropy}; \textbf{Area Under the Curve} (AUC), the area under the Receiver Operating Characteristic curve; and \textbf{Relative Improvement}, measuring the relative improvement in AUC \citep{zhou2018deep}.

Compared to baseline models, we observe a performance hierarchy across user-ad interaction types: vanilla < early interaction < late interaction < all-to-all interaction. While the vanilla two-tower model is computationally efficient, its limited user-ad interaction leads to significantly lower AUC, underscoring its representational limitations. Early- and late-interaction models (e.g., DAT, MVKE, IntTower) offer improvements but still fall short in capturing the multi-faceted nature of user interests and ad attributes. All-to-all interaction models demonstrate stronger performance due to their end-to-end interaction modeling; however, they lack pre-computability, resulting in substantial inference overhead and limited industrial applicability.

The proposed HIT model outperforms all baselines across datasets in both AUC and CEloss, indicating superior discriminative ability for user-ad matching. This advantage arises from two key components: (1) the generator, which effectively bridges static features and high-level embeddings, enabling HIT to surpass even complex models such as DCN and AutoInt with lower computational cost; and (2) the multi-head representer, which captures diverse user and ad characteristics through fine-grained subspace projections, thereby enhancing overall matching precision.

\subsection{Ablation Study}
To answer \textit{\textbf{Q2}}, we conduct several ablation experiments to demonstrate the impact of generators, multi-head representers, and their settings on the HIT model.

\begin{table}[!htbp]
\centering
\resizebox{\columnwidth}{!}{
\begin{tabular}{lllllll}
\hline
& \multicolumn{2}{c}{Alibaba}       & \multicolumn{2}{c}{MovieLens}     & \multicolumn{2}{c}{Amazon}        \\ \cline{2-7} 
Model                                                               & AUC             & CEloss          & AUC             & CEloss          & AUC             & CEloss          \\ \hline
HIT                                                                 & \textbf{0.7226} & \textbf{0.2104} & \textbf{0.9048} & \textbf{0.3026} & \textbf{0.8784} & \textbf{0.3225} \\ \hline
\multicolumn{7}{l}{\textbf{Panel A:} Remove generators and multi-head representers}                                                                                                       \\ \hline
w/o generators                                                      & 0.7170           & 0.2147          & 0.8955          & 0.3155          & 0.8690           & 0.3325          \\
w/o representers                                                    & 0.7166          & 0.2159          & 0.8954          & 0.3157          & 0.8689          & 0.3326          \\
w/o both                                                            & 0.6579          & 0.2292          & 0.8697          & 0.4559          & 0.8469          & 0.4313          \\ \hline
\multicolumn{7}{l}{\textbf{Panel B:} Remove target and non-target generators}                                                                                                             \\ \hline
w/o target                                                          & 0.7196          & 0.2127          & 0.9022          & 0.3129          & 0.8751          & 0.3251          \\
w/o non-target                                                      & 0.7216          & 0.213           & 0.8987          & 0.3144          & 0.8723          & 0.3286          \\
w/o both                                                            & 0.7170           & 0.2147          & 0.8955          & 0.3155          & 0.8690           & 0.3325          \\ \hline
\multicolumn{7}{l}{\textbf{Panel C:} Consider dynamic features}                                                                                                                           \\ \hline
\begin{tabular}[c]{@{}l@{}}w/ dynamic\\        feature\end{tabular} & 0.7205          & 0.2638          & 0.9012          & 0.3107          & 0.8711          & 0.3321          \\ \hline
\multicolumn{7}{l}{\textbf{Panel D:} Consider other types of generation loss}                                                                                                             \\ \hline
w/ MAE                                                              & 0.7204          & 0.2674          & 0.8945          & 0.3316          & 0.8691          & 0.3324          \\
w/ MSE                                                              & 0.7188          & 0.265           & 0.8933          & 0.3194          & 0.8689          & 0.3323          \\ \hline
\end{tabular}}
\caption{Main ablation experimental results. Each panel represents a set of ablation studies compared to the HIT model.}\label{tab-ablation}
\end{table}

\textit{\textbf{Q2.1.}} \textbf{What is the impact of removing generators or multi-head representers?} We remove either the generators, the multi-head representers, or both, to validate their impacts on the HIT model. Panel A in Table \ref{tab-ablation} reports the performances. The results show that removing either component, generators or representers, leads to a significant decline in model performance. Removing both components causes the model's accuracy to plummet. Hence, both the generators and the multi-head representers are essential and cannot be omitted.

\textit{\textbf{Q2.2.}} \textbf{What is the impact of generator settings on the HIT model?} One of the critical settings in the generator is the separation of the target and non-target user/ad information when handling positive and negative samples, which contributes to different informational gains. In Panel B in Table \ref{tab-ablation}, we present how such settings impact the model performances. As expected, the HIT model, including both target and non-target generators, performs the best while the one excluding them performs the worst across all datasets. This indicates that the positive and negative samples have different impacts on generators. However, their contributions can vary depending on the data pattern, as evidenced by the fact that excluding only the target generator performs better on the MovieLens and Amazon datasets. In contrast, excluding only the non-target generator performs better on the Alibaba dataset.

\textit{\textbf{Q2.3.}} \textbf{Does dynamic features help increase HIT performances?} As shown in Figure \ref{hit_model} and Eq.\eqref{eq:r_m_equation}, the generators' input only accounts for users/ads static features. This helps the generator capture long-term characteristics of users/ads and leave out short-term interest and attribute changes when mimicking the opposite towers. We have tried to include the dynamic features and report the results in Panel C in Table \ref{tab-ablation}. We can observe that including dynamic features decreases the model's performance. This stems from the fact that the dynamic features introduce extra noise when mimicking user and ad representations, thereby introducing erroneous information into the DNN module during feature crossing.

\textit{\textbf{Q2.4.}} \textbf{Can other distance metrics replace the cosine similarity in the proposed generation loss?} We replace the measurement of the distance in Eqs.\eqref{eq:loss_generation} by mean square error (MSE) and mean absolute error (MAE). Panel D in Table \ref{tab-ablation} reports the model performances. We can observe that the HIT model with cosine achieves the best performance on all datasets. Cosine similarity is capable of handling high-dimensional user and ad representations because it addresses the direction rather than the Euclidean distances. In contrast, the metrics MSE and MAE account for the straight distance between two points in space, determining geometric proximity between vectors, so they are more suitable for low-dimensional representations.

\begin{figure}[!htbp]
\centering
\begin{subfigure}[b]{0.45\columnwidth}
    \includegraphics[trim=0 45 0 20 clip, width=\columnwidth]{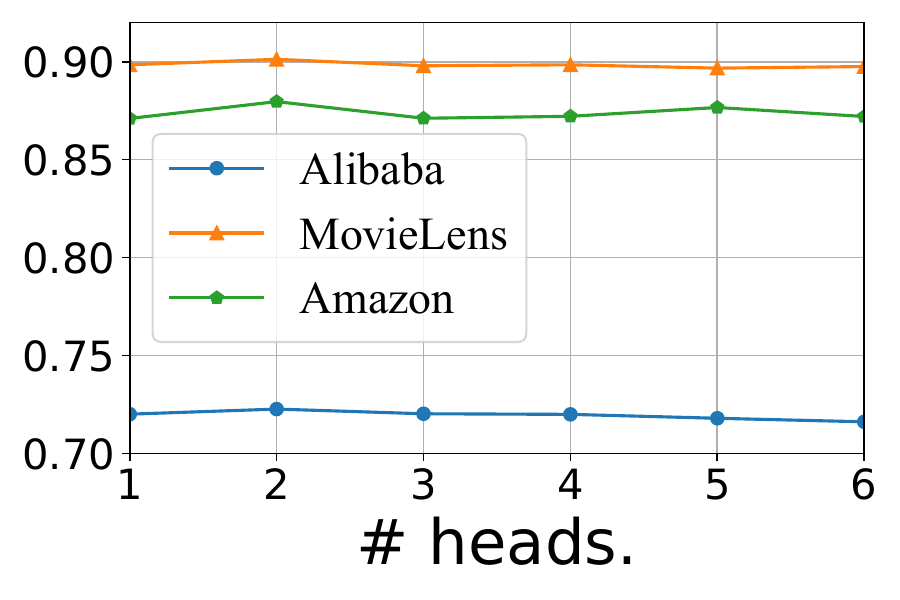}
    \caption{AUC concerning $J$.}
    \label{fig:head1}
\end{subfigure}
\hfill
\begin{subfigure}[b]{0.45\columnwidth}
    \includegraphics[trim=0 45 0 20 clip, width=\columnwidth]{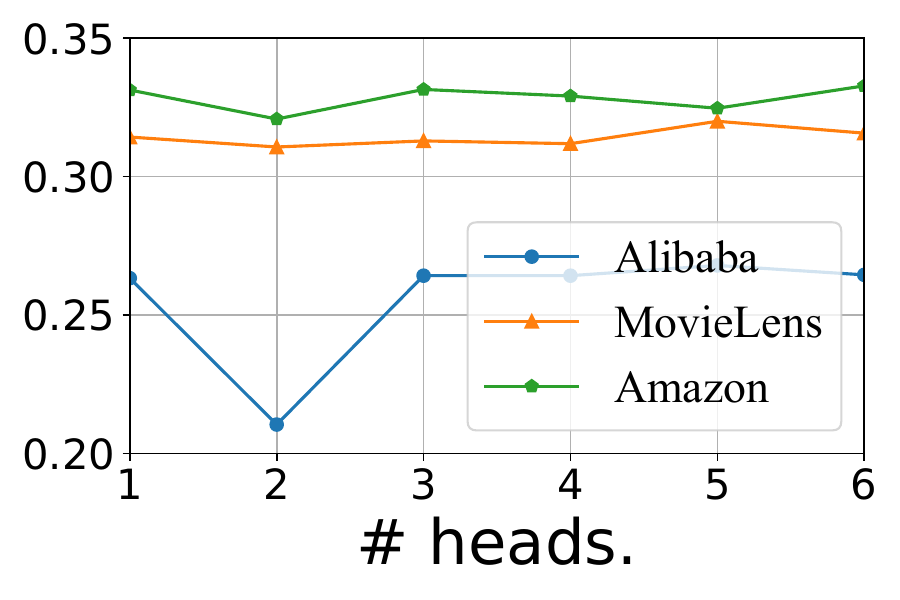}
    \caption{CEloss concerning $J$.}
    \label{fig:head2}
\end{subfigure}
\caption{Evaluation metrics concerning $J$.}
\label{fig:Results of different head numbers in multi-head representer}
\end{figure}

\textit{\textbf{Q2.5.}} \textbf{What is the best number of heads in our multi-head representer?} A critical setting for capturing fine-grained semantic relationships between user and ad embeddings is the number of heads $J$. We have tuned different values ranging from 1 to 6 on all datasets, and plotted the curves between the evaluation metrics and $J$ in Figure \ref{fig:Results of different head numbers in multi-head representer}. The results show that smaller numbers of heads tend to achieve better performances than greater $J$. The performance obtains the best when $J=2$, which is the setting in the proposed HIT model. An excessive number of heads may lead to a complex model structure and suffer the risk of overfitting, which reduces the model's generalization capability. In practice, we recommend a small $J$, but it can increase given the complexity of the data at hand.

\subsection{Investigation on Interaction Mechanism}

To answer \textit{\textbf{Q3}}, we investigate the details of how the proposed interaction mechanism works in practice. We randomly select one user (ad) and all ads (users) related to the selected one in the MovieLens test set to study both the user and ad towers mechanisms.

\begin{figure}[htbp]
    \begin{subfigure}{0.49\columnwidth}
        \centering
        \includegraphics[trim=0 45 0 20 clip, width=\textwidth]{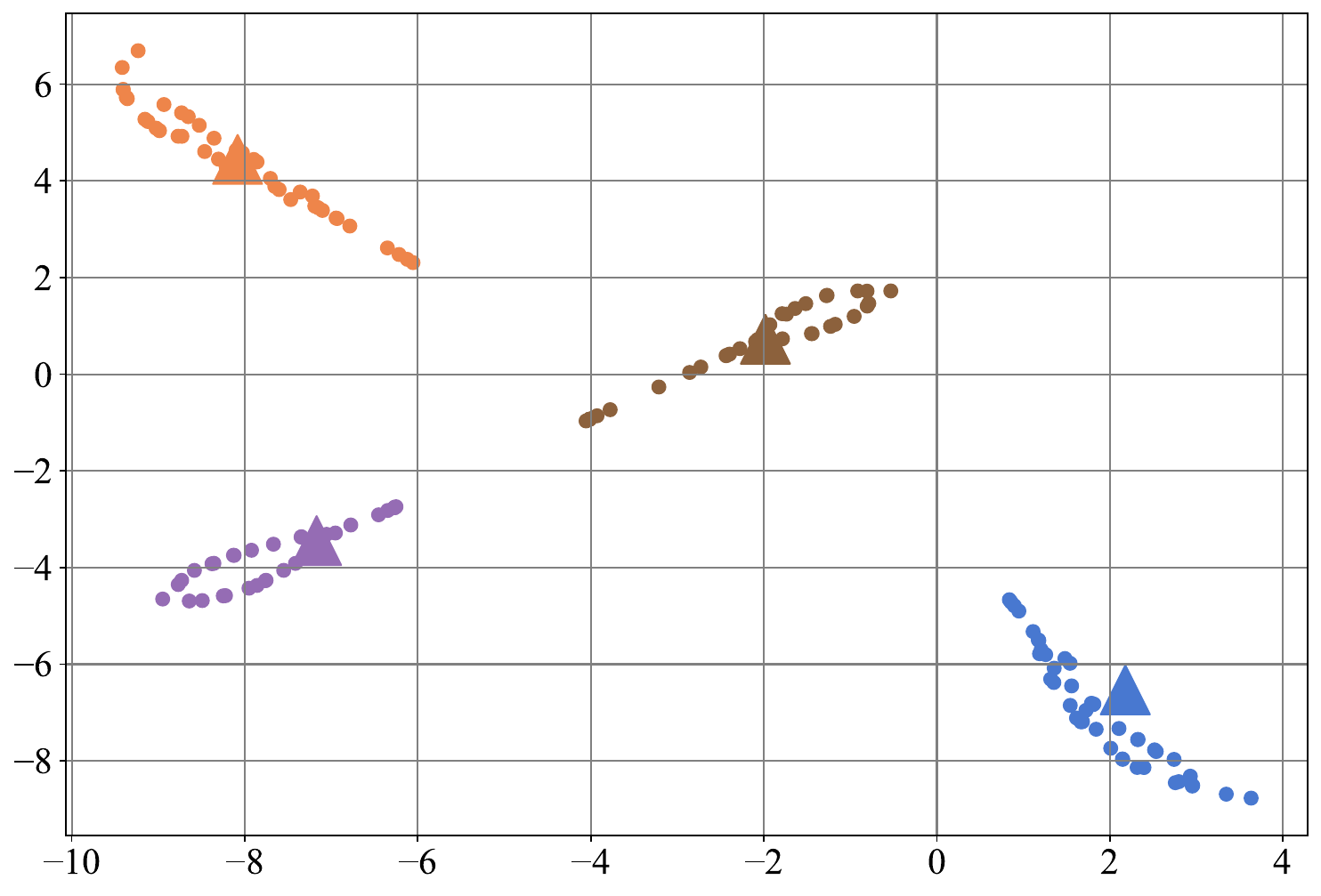}
        \subcaption{User tower. \label{fig-user-generator}}
    \end{subfigure}
    \hfill
    \begin{subfigure}{0.49\columnwidth}
        \centering
        \includegraphics[trim=0 45 0 20 clip, width=\textwidth]{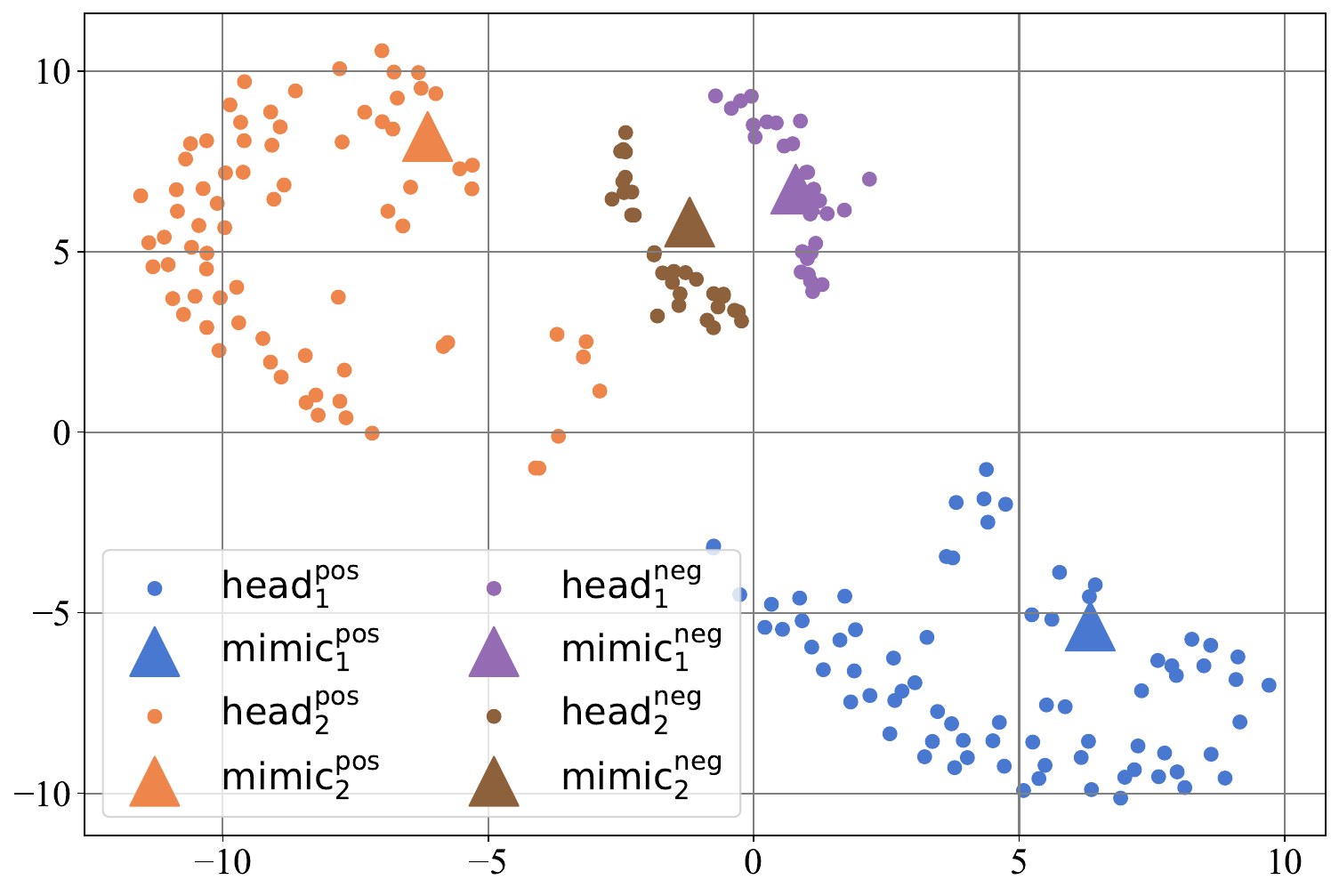}
        \subcaption{Ad tower. \label{fig-ad-generator}}
    \end{subfigure}
    \caption{Visualization of mimic/original representations.}
    \label{mechanism}
\end{figure}

The mimic vectors $\mathbf{r^m_\mathit{k}}$ produced by the generator and the multi-head representations $\mathbf{r}_{u, \mathit{(j)}}$/$\mathbf{r}_{a, \mathit{(j)}}$ from the representer are visualized via t-SNE \citep{van2008visualizing} in Figure~\ref{mechanism}. Circles denote high-dimensional outputs from the representer, while triangles denote mimic vectors generated by the generator.

Colors indicate different sample types and attention heads. In Figure~\ref{fig-user-generator}, $\mathbf{r}_{u, (1)}$ and $\mathbf{r}_{u, (2)}$—denoted as $head^{pos}_1$ and $head^{pos}_2$—represent user-favored (positive) ads, whereas $head^{neg}_1$ and $head^{neg}_2$ correspond to representations of negative samples. The generator's mimic vectors are labeled as $mimic^{pos}_1$, $mimic^{pos}_2$, $mimic^{neg}_1$, and $mimic^{neg}_2$, derived from generators trained on positive and negative samples, respectively. Figure~\ref{fig-ad-generator} follows a similar notation.

As shown in Figure~\ref{mechanism}, the multi-head representer forms coherent clusters for semantically similar samples while maintaining clear inter-group separation. This indicates its ability to project $\mathbf{h^{i}_\mathit{u}}$ and $\mathbf{h^{i}_\mathit{a}}$ into distinct semantic subspaces, effectively capturing the diverse attributes of users and ads through its multi-perspective representation.

Notably, the generator’s mimic vectors $\mathbf{r^m_\mathit{k}}$ tend to lie near the centers of their respective clusters, serving as representative proxies. This centrality suggests that the generator effectively captures the shared characteristics within clusters, indicating successful alignment between static features and high-level embeddings via learned latent correlations.

\subsection{Online A/B Test}

To address \textit{\textbf{Q4}}, we conduct rigorous online A/B testing on a leading commercial display advertising platform. The proposed HIT model is trained on a large-scale dataset comprising over 3.6 billion samples, involving more than 1 billion users and 10 million ad candidates. In the online environment, the HIT model selects several hundred ads per user for forwarding to the downstream modules illustrated in Figure~\ref{ad_flow}. These outputs form the treatment group. For the control group, we deploy the MVKE model, which is selected as the strongest-performing baseline on our private data. The two groups' subsequent decision-making processes remain identical to ensure a fair comparison. The A/B test is conducted over five days.


Table \ref{tab-online} provides the improvement of two key business metrics, including GMV and ROI, compared to the control group. There is a 1.55\% increase in GMV, indicating that the treatment group brings more revenue for the advertisers than the control group. An increase in ROI means that additional revenue does not require extra costs. The online experimental results demonstrate that the proposed HIT model can better approximate the user-ad pairs and help the following decisions to be more effective. 
\begin{table}[!htbp]
    \centering
    \begin{tabular}{lcc}
    \hline
         & GMV & ROI \\
         \hline
       Improvement  & +1.66\% & +1.55\% \\ 
       \hline
    \end{tabular}
    \caption{Relative improvement compared to control group.}
    \label{tab-online}
\end{table}

\begin{figure}[t]
\centering
\begin{subfigure}{0.23\textwidth}
    \centering
    \includegraphics[trim=100 40 100 10, scale=0.2]{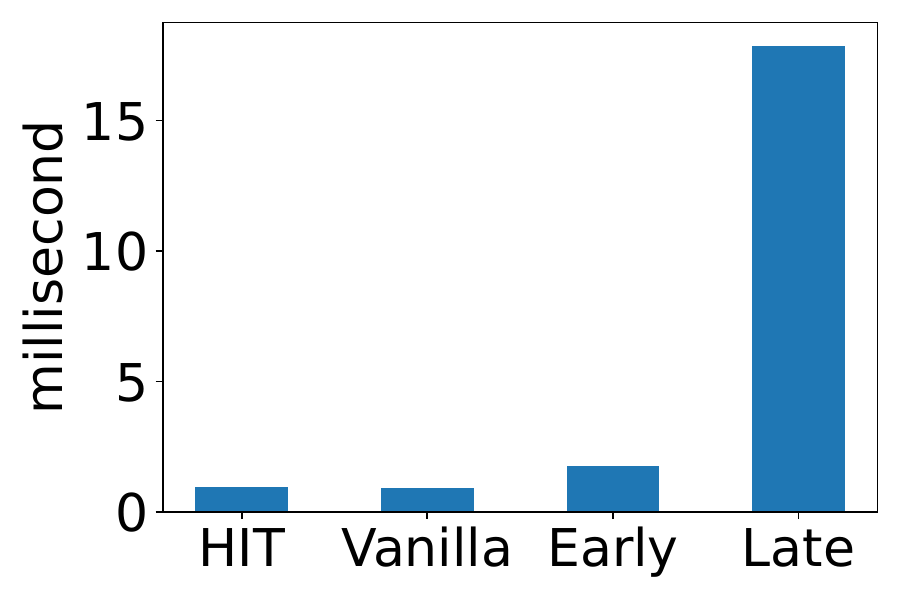}
    \subcaption{Average response time. \label{fig-QPS35000MS}}
\end{subfigure}
\hfill
\begin{subfigure}{0.23\textwidth}
    \centering
    \includegraphics[trim=100 40 100 10, scale=0.2]{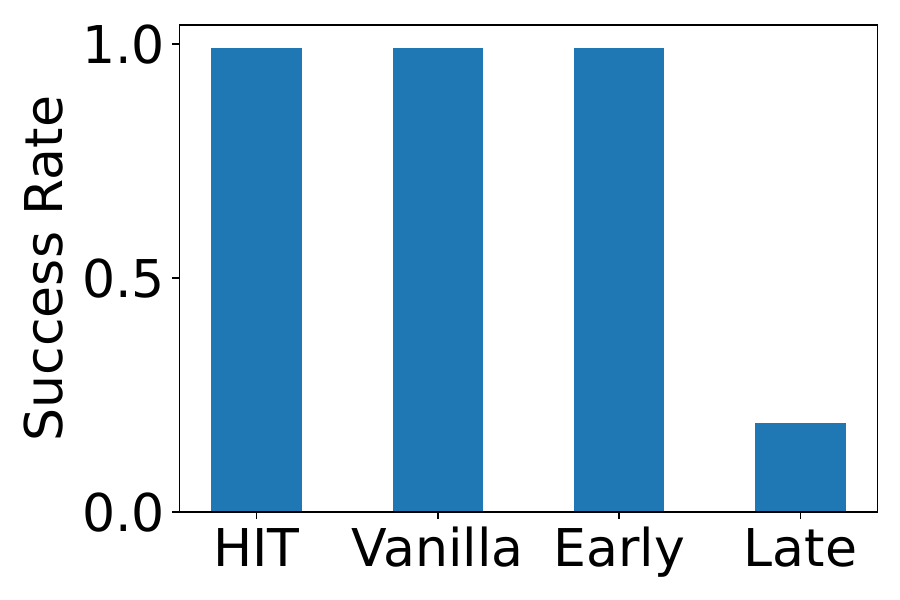}
    \subcaption{Response success rate.\label{fig-QPS35000ST}}
\end{subfigure}
\caption{Results of online efficiency. \textit{Millisecond} and \textit{Success Rate} represent the inference time and the success rate of responses under QPS=35,000.}
\label{all_qps}
\end{figure}

Figure \ref{all_qps} plots the response time and success rate concerning Queries Per Second (QPS) to examine the efficiency. Given the same QPS, the model with a smaller response time and higher success rate is more efficient. We exclude the comparison to all-to-all interaction models since they cannot sustain the QPS=35,000 (A QPS of 35,000 is the threshold metric for online services; if this threshold is not met, online deployment is not feasible). We can see that the HIT model's efficiency is very close to that of the vanilla two-tower models. The efficiency can be summarized as follows: vanilla $>$ early/HIT $>$ late $\gg$ all-to-all models.

\section{Conclusion}

We addressed the \textit{efficient-effectiveness} challenge in pre-ranking systems for online display advertising by introducing the Hierarchical Interaction-enhanced Two-tower (HIT) model. 
HIT integrates a \textit{generator} that identifies latent correlations between static features and high-level embeddings to enhance coarse-grained user-ad interactions through vector generation, while employing a \textit{multi-head representer} that models multi-faceted user/ad attributes via latent subspace projections for fine-grained matching with improved scoring precision.
Extensive experiments demonstrate that HIT consistently outperforms state-of-the-art baselines in predictive performance. Real-world deployment on the Tencent advertising platform yielded substantial business gains, including a 1.66\% increase in GMV and a 1.55\% improvement in ROI, validating HIT’s effectiveness and practical scalability in industrial settings.

\balance
\bibliographystyle{ACM-Reference-Format}
\bibliography{references}

\end{document}